\documentclass[doublecol]{epl2} 

\usepackage{color}
\usepackage{graphicx}
\usepackage{wrapfig,enumerate,slashed}

\usepackage[switch]{lineno}
\usepackage{footmisc}
\usepackage{amsmath}
\usepackage{wasysym} 
\usepackage{graphicx}
\usepackage{color}
\usepackage{comment}
\usepackage{hyperref}
\usepackage{caption}
\usepackage{subcaption}
\usepackage{amssymb}
\usepackage{adjustbox}

\newcommand{\be}{\begin{eqnarray}}
\newcommand{\ee}{\end{eqnarray}}

\newcommand{\bee}{\begin{eqnarray}}
\newcommand{\eee}{\end{eqnarray}}
\newcommand{\beeq}{\begin{equation}}
\newcommand{\eeeq}{\end{equation}}

\title{Constraining strongly coupled new physics from cosmic rays with machine learning techniques}

\author{Peter Schichtel\inst{1,2}\thanks{E-mail: \email{peter.schichtel@dfki.de; peter.schichtel@iav.de}}\
 \and Michael Spannowsky\inst{3}\thanks{E-mail: \email{michael.spannowsky@durham.ac.uk}} \and Philip Waite\inst{3}\thanks{E-mail: \email{p.a.waite@durham.ac.uk}}}
 
\shortauthor{P. Schichtel, M. Spannowsky and P. Waite}

\institute{
\inst{1} German Research Center for Artificial Intelligence (DFKI), 67663 Kaiserslautern, Germany \\
\inst{2} Ingenieurgesellschaft Auto und Verkehr (IAV), 67663 Kaiserslautern, Germany \\
\inst{3} Institute for Particle Physics Phenomenology, Department of Physics, Durham University, Durham, DH1 3LE, UK
}

\abstract{~Cosmic rays interacting with the atmosphere allow for the probing of fundamental interactions at ultra-high energies. We thus obtain limits on strongly coupled new physics models via their imprints on cosmic-ray air showers. Using the Monte Carlo event generators Herwig and HERBVI, and the air shower simulator CORSIKA, to simulate such processes, we apply machine learning algorithms to the simulated observables to discriminate the events arising via new physics from the QCD background. We then use the signal and background discrimination performance to set potential limits on the cross sections of the new physics models.}

\begin{document}

\maketitle

%%%%%%%%%%%%%%%%%%%%%%%%%%%%%%%%%%%%%%%%%%%%%%%%%%%%%
%  new section
%%%%%%%%%%%%%%%%%%%%%%%%%%%%%%%%%%%%%%%%%%%%%%%%%%%%%

\section{Introduction}
\label{sec:intro}
The recent discovery of the Higgs boson~\cite{Aad:2012tfa,
  Chatrchyan:2012xdj} was the last missing piece to establish the
Standard Model of particle physics as an effective theory describing
interactions at $\mathcal{O}(1)$~TeV, thereby confirming the paradigm that nature can be described to a high precision with perturbative quantum field theory in such an energy range. However, many UV completions of the Standard Model predict fundamental modifications to that paradigm. In particular, they predict that the theory transitions from a weakly-coupled into a strongly coupled regime not too far beyond the electroweak scale, e.g. in the range $10-100$~TeV. Examples of such theories\footnote{See also Ref.~\cite{Arkani-Hamed:2015vfh} for selected resonance cross sections and simplified models with mediators to strongly coupled sectors \cite{Englert:2016knz,Becciolini:2014lya} at 100 TeV proton-proton collisions.} are composite Higgs models \cite{Kaplan:1983sm,Caracciolo:2012je, Barnard:2013zea,Ferretti:2014qta}, little string theories \cite{Antoniadis:2011qw}, Higgsplosion \cite{Khoze:2017tjt,Khoze:2017lft,Khoze:2017uga} and classicalization \cite{Dvali:2010jz, Dvali:2012mx}. 

While the former results in the production of strongly coupled resonances (such as $Z^\prime$ or heavy scalar particles, which are usually short-lived and decay into a small number of Standard Model particles), the latter two examples result in the production of a multi-particle final state where the energy of the phenomenon is subsequently distributed over a plethora of particles, not unlike the $(B+L)$-violating sphaleron process of the Standard Model. If such processes can be realised with appreciable probabilities, separating signals with a small number of final state objects from large QCD-induced Standard Model backgrounds is a significantly bigger task in a collider environment than for final states with $\mathcal{O}(100)$ particles.

To access energies of $\mathcal{O}(10)$ TeV in fundamental interactions, protons have to be collided at $\mathcal{O}(100)$ TeV center-of-mass energies to account for the fact that the individual quarks and gluons in the proton only carry a fraction of the proton's energy. In the absence of a proton-proton collider that can access such energies, we instead focus on ultra-high-energy cosmic rays to study whether strongly coupled new physics can be probed in their interactions with the atmosphere. When a highly energetic proton hits the atmosphere, large momentum transfers occur which eventually give rise to an extended air shower of photons, hadrons and leptons. As a whole, this air shower is a highly complex object which can arguably obfuscate the hard process that initiates the shower. 

In recent years, however, for high-energy events at the LHC, novel analysis techniques have been devised to study jets (complex collimated sprays of hadrons) and their substructure \cite{Marzani:2019hun}. The remarkable success of these techniques, e.g. in discriminating electroweak scale resonances from QCD-induced backgrounds, makes it plausible that one can apply similar techniques to the study of cosmic-ray air showers in separating Standard Model processes from decays of heavy resonances or multi-particle phenomena \cite{Brooijmans:2016lfv, Jho:2018dvt}. Previous work aimed at setting limits on new physics using cosmic-ray
interactions has either predominantly focused on exploiting 
primary and secondary neutrinos~\cite{Morris:1993wg, Ringwald:2001vk, Fodor:2003bn,Illana:2004qc}, hadronic shower particles~\cite{Illana:2006xg}, or very light resonances \cite{Yin:2009yt}. Here, instead, we study whether the detailed interactions of the hard process involving very heavy particles could leave an imprint strong enough to discriminate new physics from Standard Model QCD-induced backgrounds as measured at the Pierre Auger Observatory. 

\begin{figure*}[tbh]
\centering
\begin{subfigure}[b]{0.45\textwidth}
\centering
\includegraphics[width=0.92\textwidth]{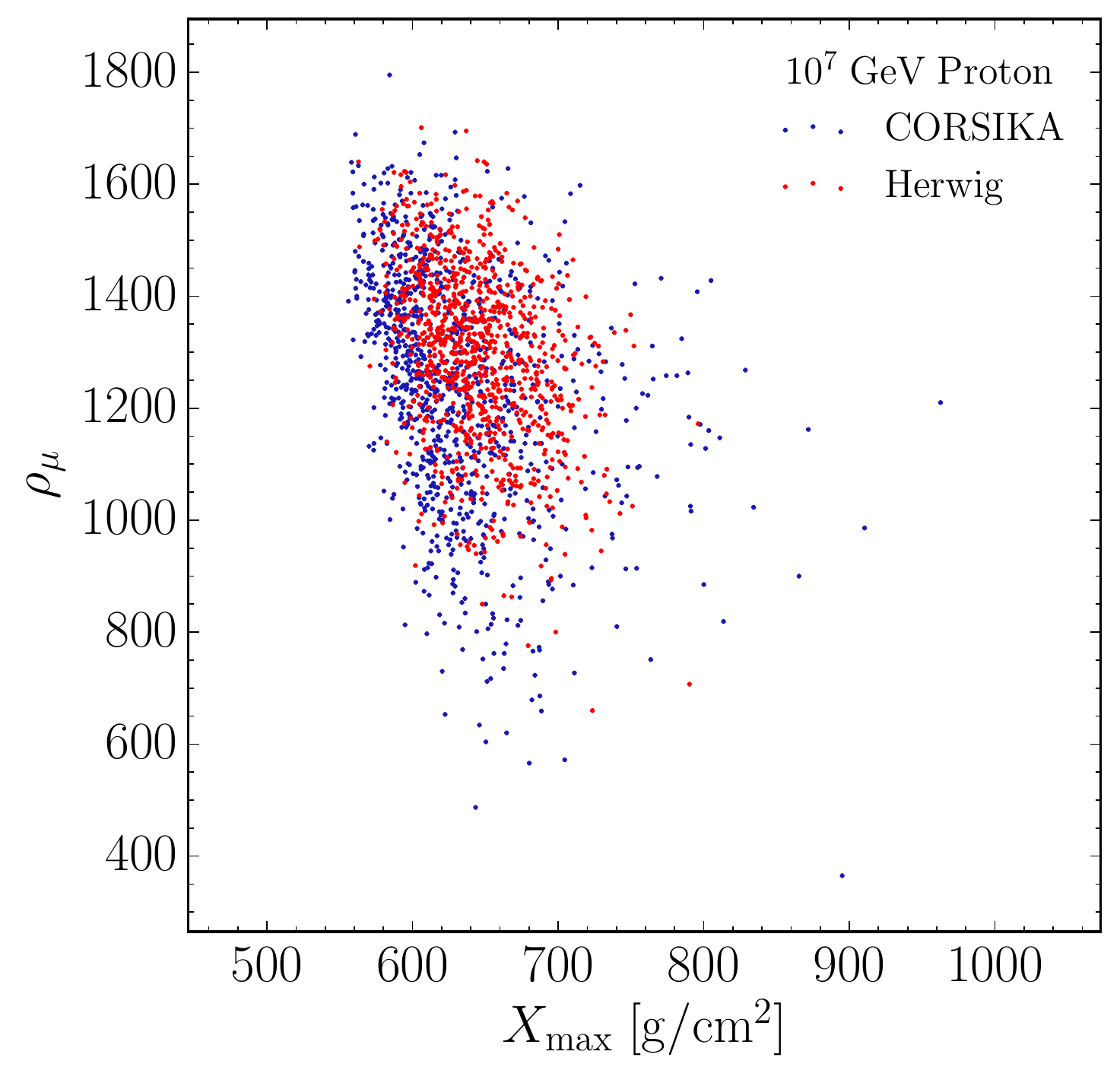}
\label{fig:corsika_vs_herwig_1e7}
\end{subfigure}
\hfill
\begin{subfigure}[b]{0.45\textwidth}
\centering
\includegraphics[width=0.92\textwidth]{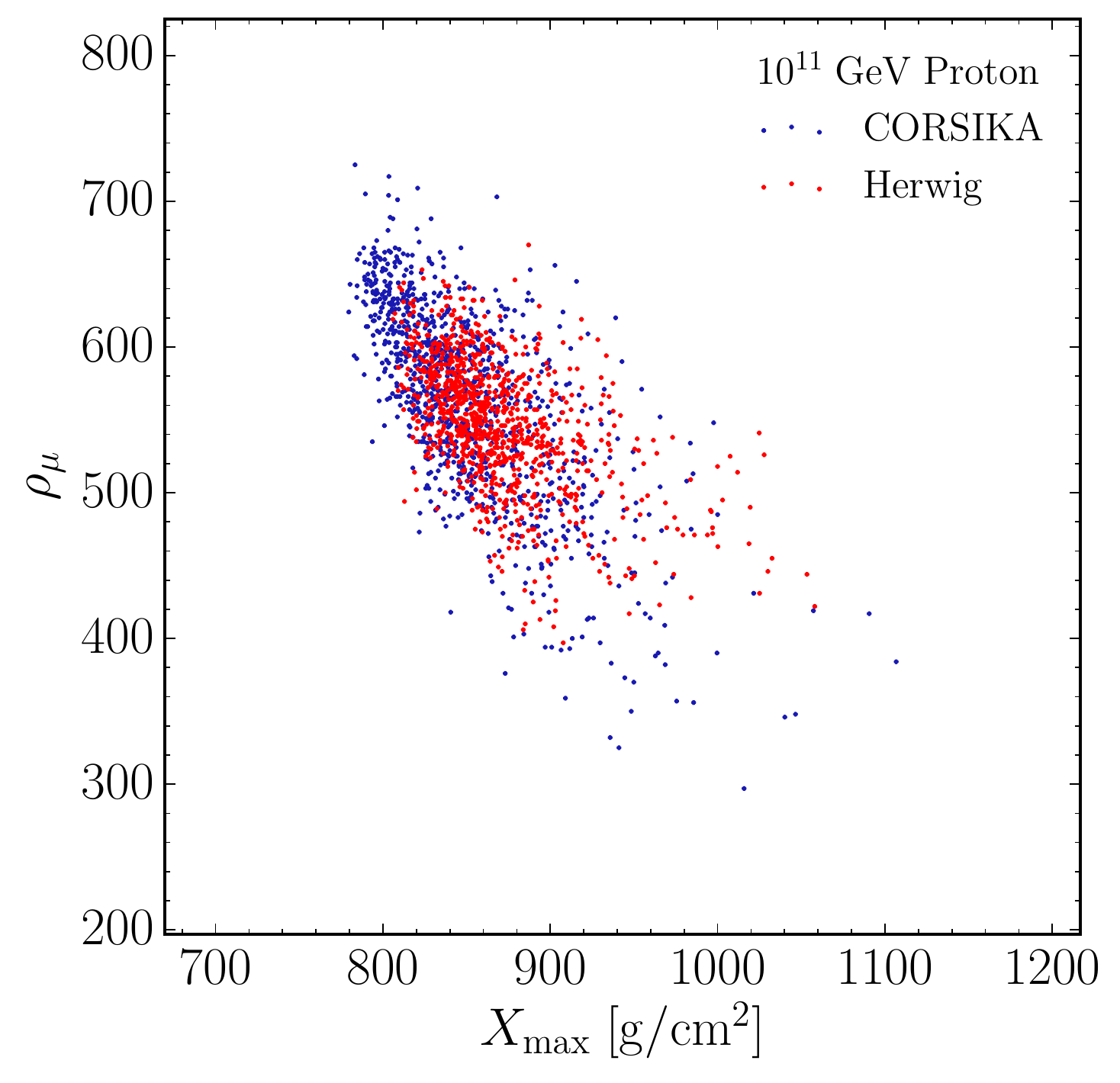}
\label{fig:corsika_vs_herwig_1e11}
\end{subfigure}
\hfill
\caption{Comparison of the two approaches of using Herwig or CORSIKA for simulating the hard process for a cosmic-ray proton at energies of (left) $10^7$ GeV and (right) $10^{11}$ GeV through their effect on the observables $X_\text{max}$ and $\rho_\mu$.}
\label{fig:corsika_vs_herwig}
\end{figure*} 

Thus, we use machine learning techniques to analyse the structure of air showers to discriminate the kinematic distributions that heavy resonances would leave compared to QCD-induced processes. 
First, we describe the simulation setup, where we use Herwig and HERBVI to generate the hard processes, followed by the simulation of the air shower using CORSIKA. Then, we show the effects of the new physics models on two air shower observables compared to the background QCD process. Finally, we train machine learning algorithms to classify the events and use this to derive simple estimates of the limits on the cross sections of these processes.

\section{Simulation Setup}
\label{sec:simulation}

In this section we describe all the steps in our simulation of cosmic-ray air showers from models of new physics.

\subsection{New Physics Processes}

To represent possible processes that can arise in non-perturbative solutions to, and UV completions of, the Standard Model, we consider a $(B+L)$-violating sphaleron process, a heavy gauge boson $Z^\prime$ decaying to two Standard Model photons, and a heavy scalar boson $h^\prime$ decaying to two Standard Model leptons. The masses of the $Z^\prime$ and $h^\prime$ resonances are 10 TeV, with widths of 100 GeV.

The sphaleron process we study includes a change in baryon and lepton numbers of $\Delta B = \Delta L = -3$ and is of the form $qq \to 7 \bar{q} + 3 \bar{l} + n_V W/Z + n_h h$, where $n_V$ and $n_h$ are the numbers of electroweak gauge bosons and Higgs bosons, respectively. Since it was suggested in Refs.~\cite{Ringwald:1989ee,Khoze:1990bm,Khoze:1991mx,Tye:2015tva} that the production cross section for sphalerons is enhanced if produced in association with many gauge bosons, in our simulation we select $n_V = 24$ and $n_h=0$. Such sphalerons could also be searched for at IceCube \cite{Ellis:2016dgb} or at high-energy proton-proton colliders \cite{Ellis:2016ast,Ringwald:2018gpv}, and if observable, they could improve our understanding of the underlying mechanism of electroweak symmetry breaking \cite{Spannowsky:2016ile}.
At the level of observability of a high-energy collision on the surface of our atmosphere, such a multi-particle production process mimics the kinematic features induced by processes from Higgsplosion or classicalization. Thus, we will take the sphaleron as representative of models with enhanced production mechanisms for elementary $2 \to n$ scatterings, where $n \gg 1$.

\subsection{Hard Interaction Simulation}

To simulate the hard interaction for the background QCD and heavy $Z^\prime$/$h^\prime$ processes, we use the Herwig 7 \cite{Bellm:2015jjp} Monte Carlo event generator. Herwig collides the two protons, computes the partonic interaction, and simulates the parton shower as well as the hadronic phase transition. 
To generate the sphaleron processes, we use the HERBVI \cite{Gibbs:1994cw,Gibbs:1995bt} tool which is implemented in Herwig. The final-state particles after hadronisation are then passed to the air shower simulation.

\subsection{Air Shower Simulation}

\begin{figure*}[!tb]
\centering
\begin{subfigure}[b]{0.45\textwidth}
\centering
\includegraphics[width=0.92\textwidth]{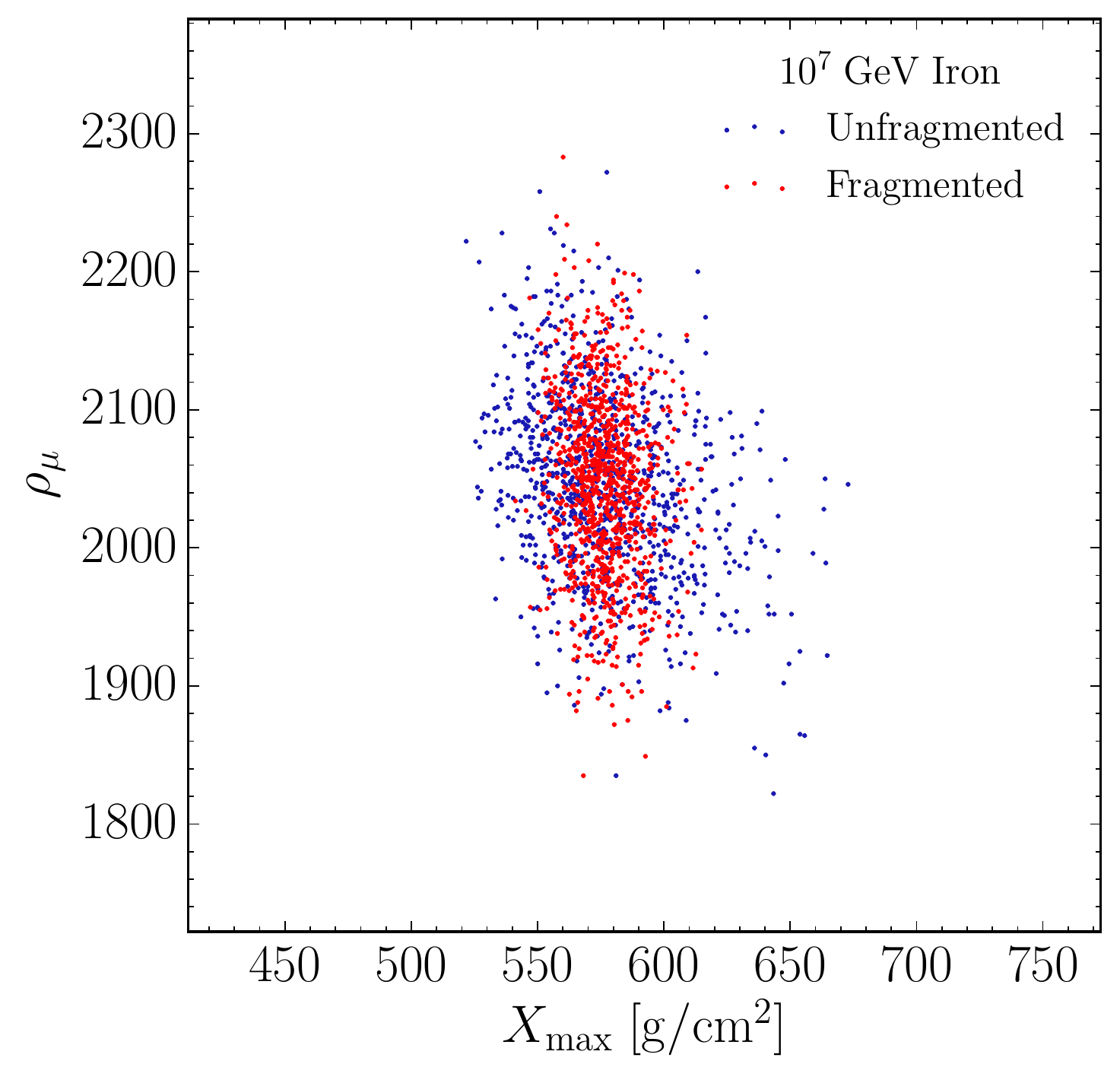}
\label{fig:fragment_comparison_1e7}
\end{subfigure}
\hfill
\begin{subfigure}[b]{0.45\textwidth}
\centering
\includegraphics[width=0.92\textwidth]{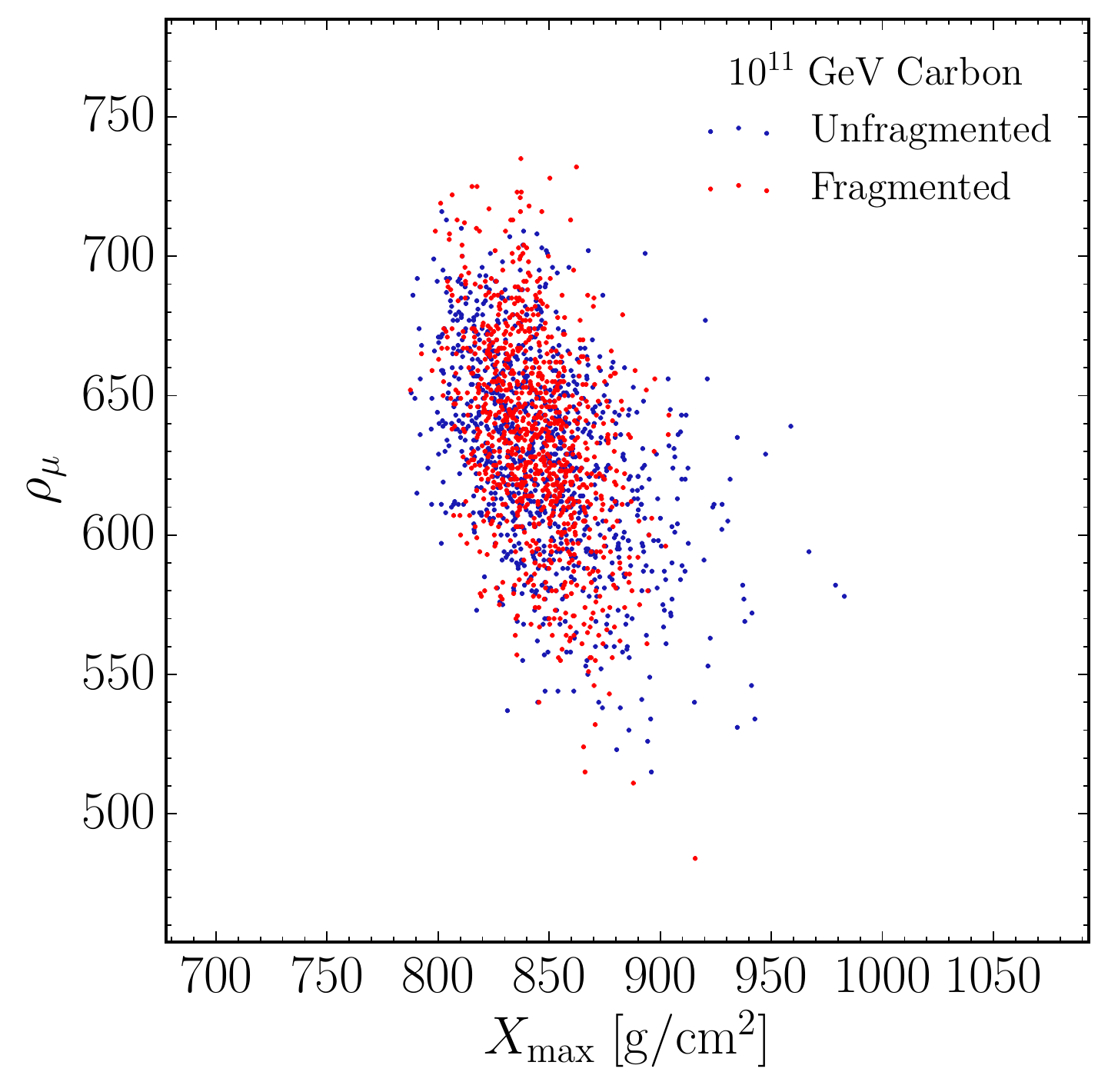}
\label{fig:fragment_comparison_1e11}
\end{subfigure}
\hfill
\caption{Comparison of the two nucleonic fragmentation models for (left) a cosmic-ray iron at $10^7$ GeV and (right) a cosmic-ray carbon at $10^{11}$ GeV through their effect on the observables $X_\text{max}$ and $\rho_\mu$.}
\label{fig:fragment_comparison}
\end{figure*} 

A cosmic-ray air shower is the phenomenon of observable secondary particles produced by a high-energy cosmic ray colliding with the upper atmosphere. In the following we briefly describe the different stages of such a shower.

The process starts with a cosmic ray heading towards the Earth, which we call the primary particle. In principle any particle could be the primary particle in the collision. However, in this work we focus on nuclear matter, and as representatives of the table of elements we choose a proton, carbon and iron.

Usually ordinary high-energy QCD describes the hard interaction when a primary particle hits an air nucleus in the upper atmosphere. However, the probability for the particular process is determined by its cross section, and in this study we also consider the other processes described above for the hard interaction. Regardless of the physics guiding the hard interaction, there will be a QCD parton shower as well as a hadronic phase transition.

\begin{table}[!t]
  \renewcommand{\arraystretch}{1.5}
  \centering
  \begin{tabular}{c|cccc}
    $E_\text{lab}^\text{primary}$ [GeV] & ~$E_\text{CM}^\text{p}$ & $E_\text{CM}^\text{C}$ & $E_\text{CM}^\text{Fe}$ & [TeV]  \\
    \hline \hline
    $10^7$~ & ~4.3    & 1.3    & 0.6  &\\
    $10^8$~ & ~13.7   & 4.0    & 1.8  &\\
    $10^9$~ & ~43.3   & 12.5   & 5.8  &\\
    $10^{10}$~ & ~137.0  & 39.5   & 18.3 &\\
    $10^{11}$~ & ~433.1  & 125.0  & 57.8 &\\
    \hline
  \end{tabular}
  \caption{Centre-of-mass collision energies corresponding to the primary particle energies considered.}
  \label{tab:energies}
\end{table}

As the interaction located in the upper atmosphere is directed downwards, a cascade of secondary interactions will follow. This is the air shower. The secondary particles will produce bremsstrahlung of any form. Furthermore, they will collide with other air molecules, feeding the cascade until the total energy is diluted and the shower dies away.

In experiments like the Pierre Auger Observatory several detectors are used to capture a signal from the air shower. Firstly, there are 1660 water Cherenkov counters on the ground, which measure both muons, and the electromagnetic shower component. These detectors count high-energy muons and establish an estimate of the distribution of the muon density at ground level. Furthermore, there are fluorescence detectors \cite{Abraham:2009pm}, which measure the fluorescence emission of air molecules in the ultraviolet range as a function of atmospheric depth.

\begin{figure*}[!t]
\centering
\includegraphics[width=\textwidth]{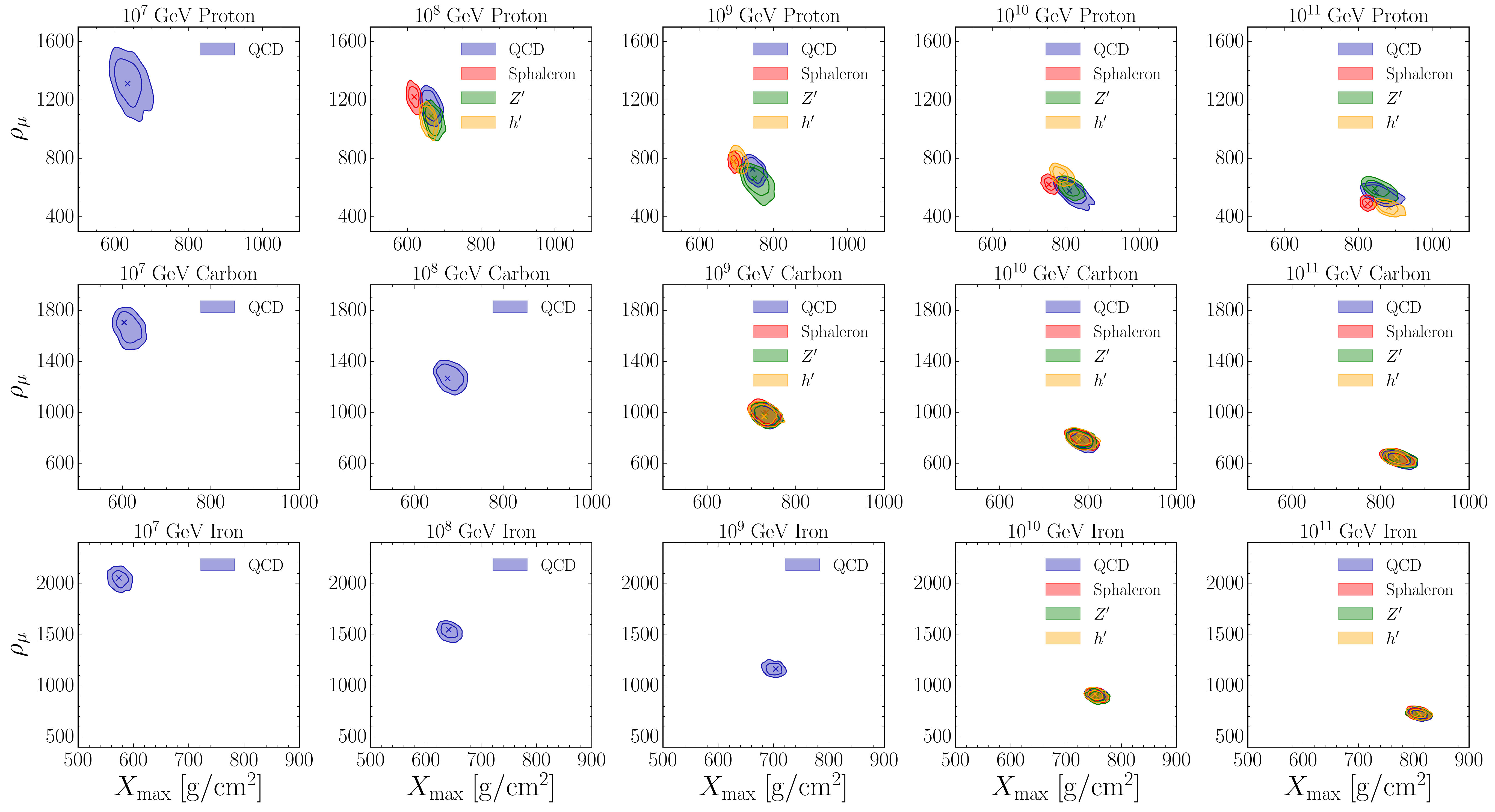}
\caption{The $X_\text{max}$ and $\rho_\mu$ distributions of the new physics models vs the QCD background for each primary particle considered. Only the new physics processes which are kinematically allowed are shown. The axis ranges are held fixed in each row of plots to show the effect of increasing the energy of each primary.}
\label{fig:New_physics_model_comparison}
\end{figure*}

To analyse new physics in cosmic-ray air showers we need to simulate the whole interaction chain described above. To do so, we process the particles generated from Herwig and HERBVI with the CORSIKA \cite{Heck:1998vt} air shower simulator. We use the GHEISHA \cite{Fesefeldt:1985yw} interaction model to treat the low-energy hadronic interactions, and the QGSJET \cite{Kalmykov:1997te} interaction model to treat high-energy hadronic interactions. A thinning procedure is applied to the shower simulation, which restricts the number of particles in each shower stage as a computational requirement.

The incoming primaries that we simulate have zero inclination and interact at a height of 18~km, with energies ranging from $E_{\mathrm{lab}}=10^7$ GeV to $E_{\mathrm{lab}}=10^{11}$ GeV. The corresponding centre-of-mass (CM) collision energies for the hard interaction, which consists of a proton in the cosmic-ray nucleus interacting with a proton in the air nucleus, are given by $\sqrt{s}\simeq\sqrt{2m_{\mathrm{p}}E_{\mathrm{lab}}/A_{\mathrm{N}}}$. Here, $A_{\mathrm{N}}$ is the atomic weight of the primary nucleus: $A_{\mathrm{N}}=1$ for a proton, $A_{\mathrm{N}}=12$ for carbon and $A_{\mathrm{N}}=56$ for iron. For the carbon and iron nuclei, the energy is assumed to be evenly distributed amongst its nucleons. Table \ref{tab:energies} shows the values of the collision energies corresponding to the primary particles that we consider.

From the simulation results, we extract the number of muons $\rho_\mu$ observed at ground level, having survived through the thinning procedure. We do not apply a dethinning procedure to this observable \cite{2012APh....35..759S}. In addition, from the distribution $N(X)$ of charged particles as a function of the shower depth $X$, we can deduce the shower maximum $X_\text{max}$ by performing a $\chi^2$-fit of a Gaisser-Hillas function \cite{1977ICRC....8..353G} to the data. This function is given by,

\begin{equation}
\label{gaisser}
N(X) = N_{\mathrm{max}}\left(\frac{X-X_0}{X_{\mathrm{max}}-X_0}\right)^{\frac{X_{\mathrm{max}}-X_0}{\lambda}}e^{\frac{X_{\mathrm{max}}-X}{\lambda}}~,
\end{equation} 
where $N_{\mathrm{max}}$, $X_{\mathrm{max}}$, $X_0$ and $\lambda$ are to be determined from the fit. In principle there is no reason why one should not include more observables usually studied in air shower experiments, such as the risetime. However, for the purposes of this study we limit it to just these two observables to determine whether these are sufficient for a meaningful discrimination between signal and background, and leave a more complicated analysis with more observables to future studies.

As a test of the reliability of using Herwig, with its capability for generating new physics processes, to generate the hard interaction and then processing the events with CORSIKA, we can also generate the full primary-to-air-shower chain for the QCD events with CORSIKA alone by using its own hard process simulation. We find that there is good agreement between them, and in Fig.~\ref{fig:corsika_vs_herwig} we show a comparison of the $\rho_\mu$ and $X_\text{max}$ distributions for the two approaches for a primary proton at both $10^7$ GeV and $10^{11}$ GeV, which spans the energy range we consider. The differences are small, although the distributions are not identical, but for the purposes of this study we will ignore any small systematic uncertainties that may arise due to the use of Herwig as the hard process generator.

Since we are not only interested in ordinary proton-proton interactions, but actually study nucleus-air collisions as well, we need to model the additional nucleonic complexity. As the air is at rest and its binding energy is low compared to the energies we are interested in, we regard it as a stationary proton. However, we cannot use such a simple ansatz for the high-energy primary particle. In principle, we might view the interaction of a nucleus with a proton in the air as a proton-proton interaction. However, we have to take the nucleonic remainder of the now-destroyed primary into account. There are two extremes we can study. We could assume that the impact was so fast that the nucleus stays untouched except with one fewer proton. On the other hand, we could assume that the nucleus is destroyed and completely fragments into its proton and neutron components. A comparison of both approaches is shown in Fig.~\ref{fig:fragment_comparison} for a cosmic-ray iron at $10^7$ GeV and a cosmic-ray carbon at $10^{11}$ GeV. We find the differences between the two extremes are small, and so for the rest of this study we consider a completely fragmented remainder nucleus.

\section{Results and Limits}
\label{sec:results}

In this section we show the effect of the new physics models on the two air shower observables, and train machine learning algorithms to classify the events into signal and background classes. From this, we derive possible limits on the cross sections of the new physics processes.

\subsection{Classification of new physics events}

\begin{figure}[!t]
\centering
\includegraphics[width=0.85\columnwidth]{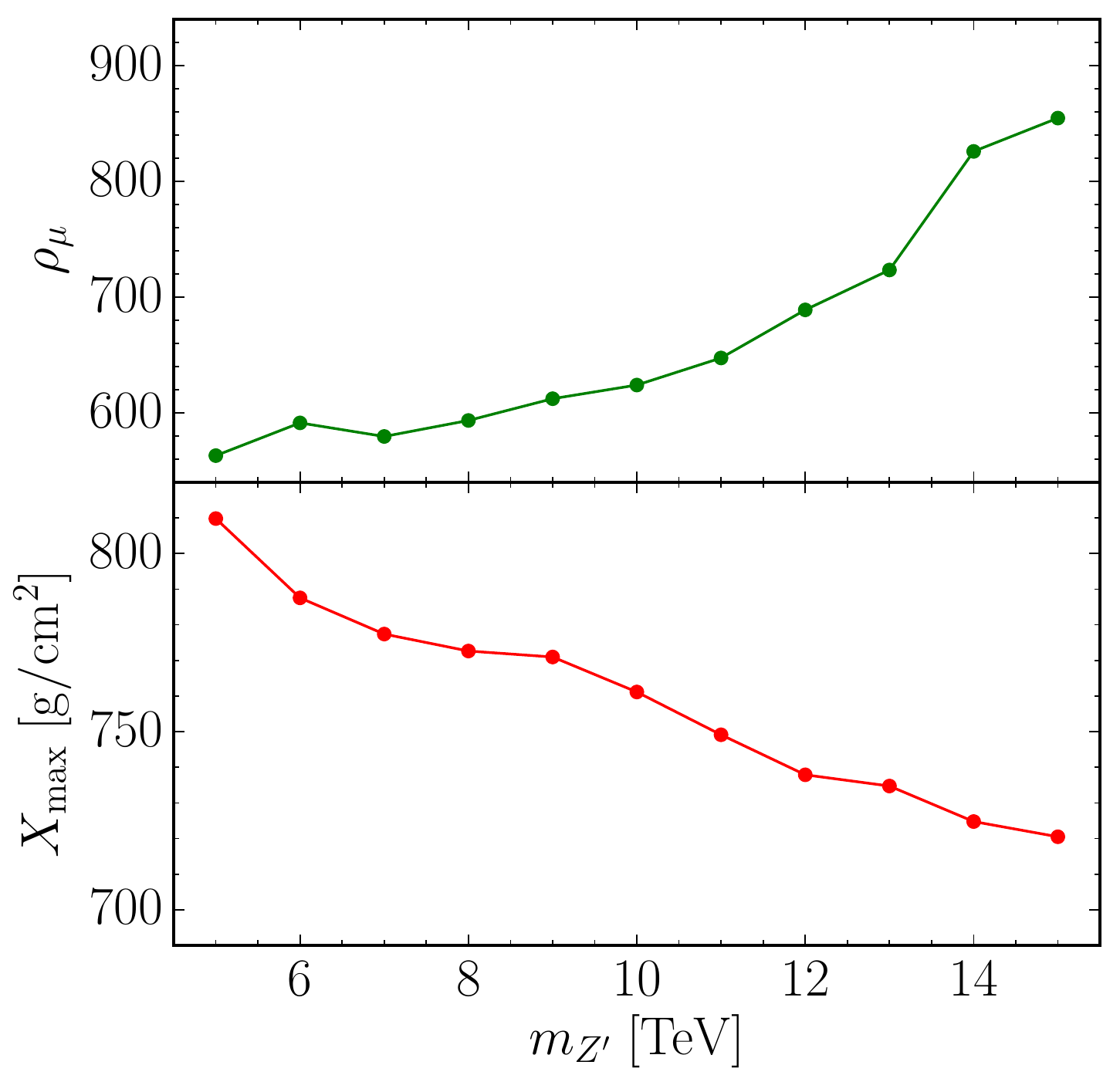}
\caption{Effect of varying the mass of the $Z^\prime$ on the average values of (lower panel) $X_\text{max}$ and (upper panel) $\rho_\mu$  for a primary proton at $10^9$~GeV.}
\label{fig:Mass_dependence}
\end{figure}

The $X_\text{max}$ and $\rho_\mu$ distributions for each new physics model in each energy and primary bin are presented in Fig.~\ref{fig:New_physics_model_comparison}, along with the background QCD distributions. The distributions shown have each been calculated from 1000 simulated points using a Gaussian kernel density estimate, with the cross showing the maximum of the distribution and the two contours enclosing 68\% and 95\% of the data. We also show the effect on the average $X_\text{max}$ and $\rho_\mu$ values as a function of the mass of the $Z^\prime$ in Fig.~\ref{fig:Mass_dependence} for a primary proton at $10^9$~GeV, and we expect this behaviour to be representative of variations in the mass for other bins and new physics models.

It is clear from the plots for carbon and iron in Fig.~\ref{fig:New_physics_model_comparison} that the new physics effects are washed out by the interactions of the remainder nucleus, and thus the parameter distributions are almost identical. Therefore, we only consider the four proton bins in the energy range $10^8-10^{11}$~GeV where the processes are kinematically possible, with the assumption that the energy and primary compositions can be determined independently of these parameters\footnote{We note that there is a relationship between $X_\text{max}$ and the composition of the primary. Indeed, one could be tempted to interpret variations of the primary composition as being potential signs of new physics. However, for the sake of this analysis we ignore these effects and their systematics, and assume that the primary compositions and energies are well-determined.}, so that these parameters can be used for the new physics classification.

\begin{figure}[!t]
\centering
\includegraphics[width=0.85\columnwidth]{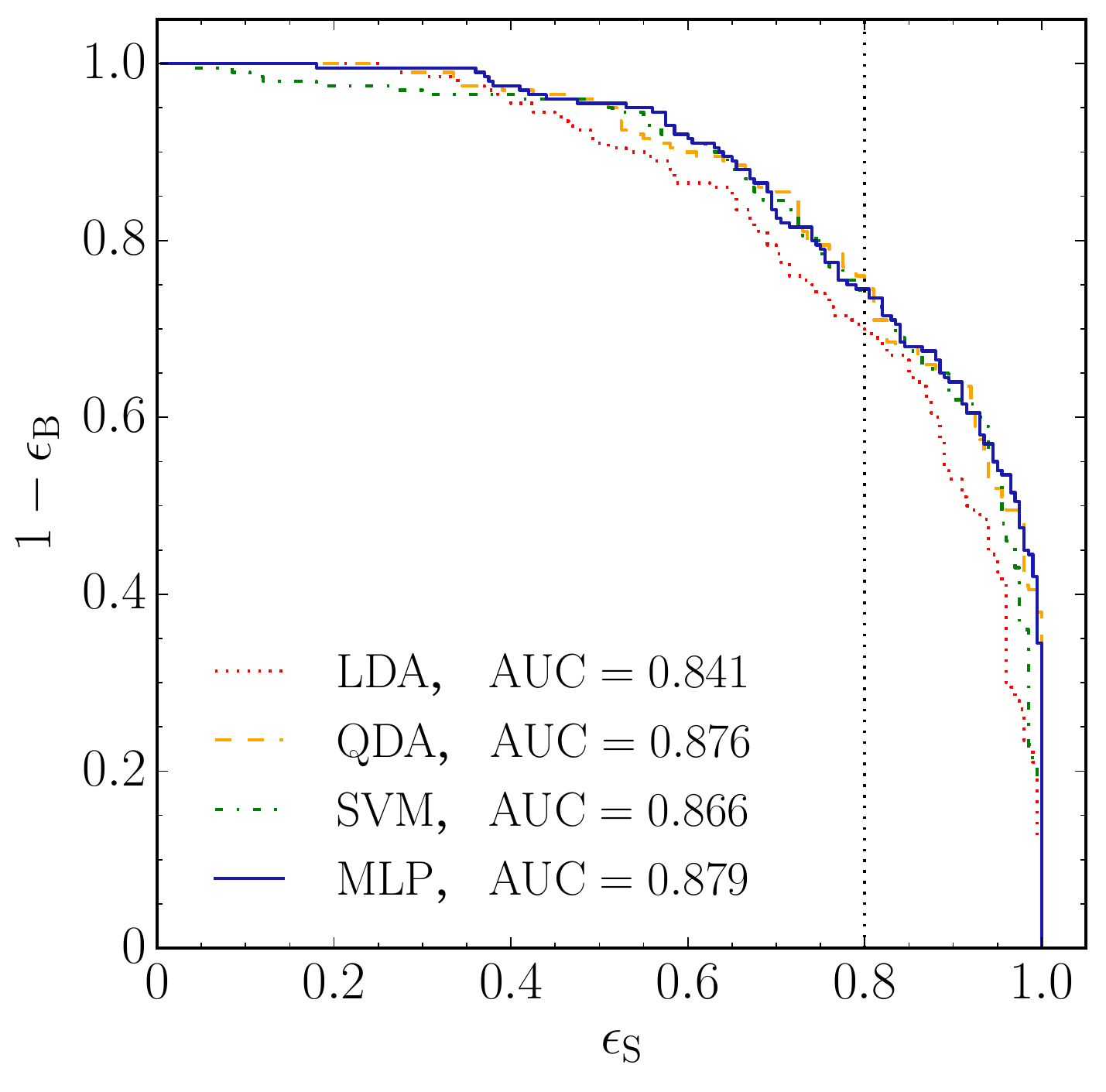}
\caption{ROC curves for the four machine learning algorithms trained to classify $Z^\prime$ vs QCD background events for a primary proton at $10^9$~GeV. The dotted line shows the chosen signal efficiency of $\epsilon_{\mathrm{S}}=0.8$.}
\label{fig:ROC_curves}
\end{figure}

In each of these energy and primary bins, we train a machine learning algorithm to independently classify the three new physics models vs the QCD background in the two-dimensional parameter space of $X_\text{max}$ and $\rho_\mu$. The machine learning algorithms that we use are a linear discriminant analysis (LDA), a quadratic discriminant analysis (QDA), a support vector machine (SVM) and a multilayer perceptron (MLP), and we use Scikit-learn \cite{Pedregosa:2012toh} for their implementation.

\begin{figure}[!t]
\centering
\includegraphics[width=0.85\columnwidth]{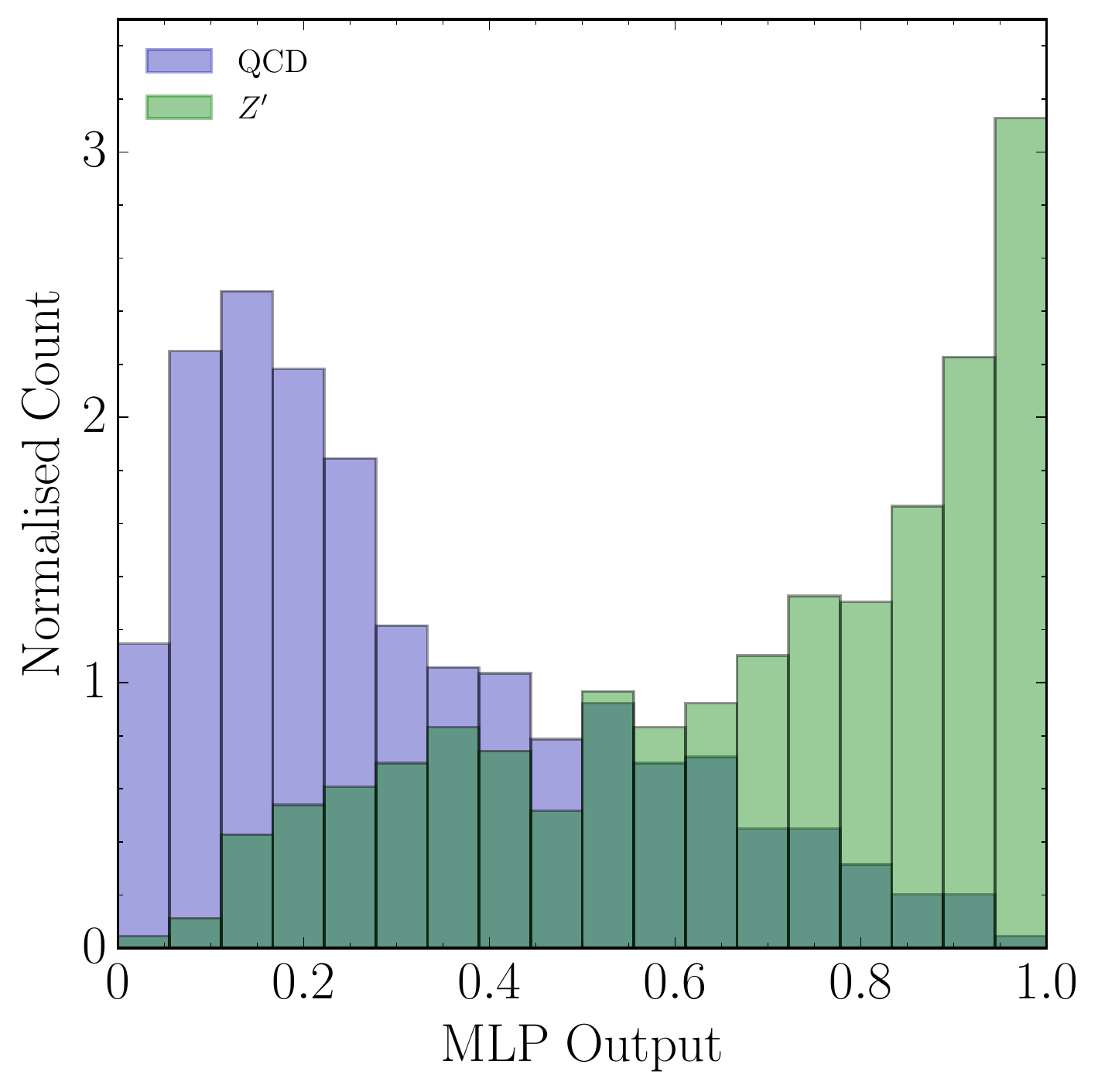}
\caption{Output of the MLP trained to classify $Z^\prime$ vs QCD background events for a primary proton at $10^9$~GeV. Larger values of the output indicate a higher confidence from the MLP model that the particular event is a signal event.}
\label{fig:MLP_output}
\end{figure}

For each new physics model and bin combination, the 2000 data points (1000 signal and 1000 background) are split into training, validation and test sets. We perform hyperparameter scans over the important hyperparameters of each algorithm, and the algorithm which has the highest accuracy on the validation set is used in order to prevent overfitting the model to the training set. We then calculate the ROC curve for each new physics model on the test set, which allows one to easily obtain the background efficiency $\epsilon_{\mathrm{B}}$ for any chosen signal efficiency $\epsilon_{\mathrm{S}}$. Fig.~\ref{fig:ROC_curves} shows the ROC curves for the $Z^\prime$ vs QCD background classification for a primary proton at $10^9$~GeV for the four machine learning algorithms considered, along with the area-under-curve (AUC) scores for each algorithm. For this particular energy and primary bin, we find that the MLP has the highest AUC score. In Fig.~\ref{fig:MLP_output}, we show the output of the MLP on the signal and background test sets, where a larger output corresponds to a higher confidence from the MLP model that the particular event is a signal event. We see that the MLP is able to discriminate most events correctly.

\subsection{Limits}

Following the analysis in Ref.~\cite{Brooijmans:2016lfv}, we can use a simple counting procedure in each proton bin to set a limit for the cross section of each new physics process in terms of the proton-air cross section. The probability for a new physics process to occur in the collision of a proton with the air can be expressed as,
\begin{equation}
\mathcal{P}_{\mathrm{new}} = A\frac{\sigma_{\mathrm{new}}}{\sigma_T(E_{\mathrm{lab}})}~,
\end{equation}
where $\sigma_T(E_{\mathrm{lab}})$ is the energy-dependent proton-air cross section, and $A=14.6$ is the average atomic mass of air. For a measured number of $N$ events, with a signal efficiency of $\epsilon_{\mathrm{S}}$ and a background efficiency of $\epsilon_{\mathrm{B}}$, we can set a 95\% confidence limit by requiring that $S/\sqrt{S+B}\gtrsim 2$, where $B=\epsilon_{\mathrm{B}}N$ and $S=\epsilon_{\mathrm{S}}N A \sigma_{\mathrm{new}}/\sigma_T$. Assuming that the number of background events is far greater than the number of signal events, this gives the limit,
\begin{equation}
\sigma_{\mathrm{new}} \lesssim \sqrt{\frac{4\epsilon_{\mathrm{B}}}{\epsilon_{\mathrm{S}}^2NA^2}}\sigma_T \equiv f \sigma_T~.
\end{equation}
The efficiencies $\epsilon_{\mathrm{S}}$ and $\epsilon_{\mathrm{B}}$ can be read off from the ROC curves. Choosing a signal efficiency of $\epsilon_{\mathrm{S}}=0.8$, the corresponding background efficiencies are shown in Table~\ref{tab:limits}, with the associated limit factor $f$ for a representative number of $N$ events in each bin, which reflects the suppression of the cosmic-ray flux as a function of energy \cite{Fenu:2017hlc,Gora:2018xty}. In cases where very strong separation is possible, the background efficiency is set to a minimum value of $\epsilon_{\mathrm{B}}=0.05$ to ensure that the limits are conservative estimates. We also show in Table~\ref{tab:limits} the derivable limits for the case where a systematic uncertainty of 5\% on the background is assumed. In this case, the 95\% confidence limit is obtained from requiring $S/\sqrt{S+B+\delta^2B^2}\gtrsim 2$, where $\delta=0.05$ is the systematic uncertainty.

\begin{table}[!t]
  \renewcommand{\arraystretch}{1.8}
  \centering
  \begin{adjustbox}{width=\columnwidth}
  \begin{tabular}{cc|cc|cc|cc}
     & & \multicolumn{2}{c|}{~Sphaleron} & \multicolumn{2}{c|}{$Z^\prime$} & \multicolumn{2}{c}{$h^\prime$}  \\
    \hline \hline
    $E_\text{lab}^\text{P}$ [GeV] & $N$ & $\epsilon_{\mathrm{B}}$ & $f$ & $\epsilon_{\mathrm{B}}$ & $f$ & $\epsilon_{\mathrm{B}}$ & $f$\\
    \hline
    $10^8$ & $50000$~ & 0.05    & {\large $^{0.00017}_{0.0046}$} & 0.28 & {\large $^{0.00041}_{0.0024}$}  & 0.14  & {\large $^{0.00029}_{0.0012}$}\\
    $10^9$ & $10000$~ & 0.05    & {\large $^{0.00038}_{0.00057}$} & 0.26 & {\large $^{0.00087}_{0.0023}$}  & 0.12  & {\large $^{0.00059}_{0.0012}$}\\
    $10^{10}$ & $1000$~ & 0.05    & {\large $^{0.0012}_{0.0013}$} & 0.60 & {\large $^{0.0042}_{0.0066}$}  & 0.05  & {\large $^{0.0012}_{0.0013}$}\\
    $10^{11}$ & $50$~ & 0.05    & {\large $^{0.0054}_{0.0054}$} & 0.31 & {\large $^{0.013}_{0.014}$}  & 0.09  & {\large $^{0.0073}_{0.0073}$}\\
    \hline
  \end{tabular}
  \end{adjustbox}
  \caption{Background efficiencies $\epsilon_{\mathrm{B}}$ and derived limit fractions $f$ for the new physics cross sections for a selected signal efficiency $\epsilon_{\mathrm{S}}=0.8$, and representative numbers of events $N$. The limit fractions $f$ are shown for (upper number) no systematic uncertainty in the background and (lower number) a 5\% systematic uncertainty in the background.}
  \label{tab:limits}
\end{table}

\begin{figure}[!t]
\centering
\includegraphics[width=0.9\columnwidth]{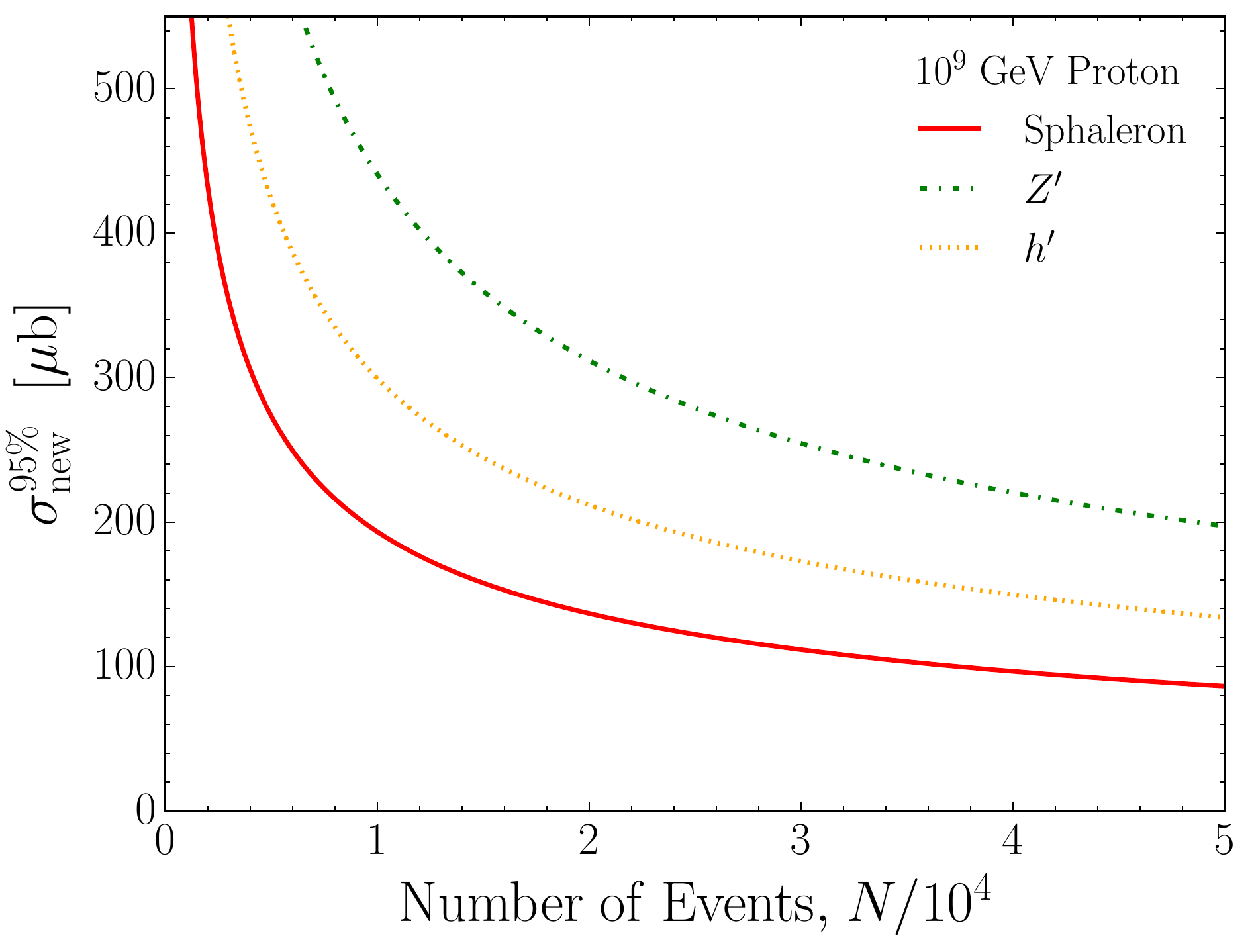}
\caption{95\% confidence limits on the cross sections of the new physics processes obtainable for a cosmic-ray proton at $10^9$~GeV, as a function of the number of events $N$ observed.}
\label{fig:cross_section_limits}
\end{figure}

For proton energies in the range $10^8-10^{11}$~GeV, the proton-air cross section ranges from $\sim 450$~mb to $600$~mb \cite{Collaboration:2012wt}. Thus the limits on the new physics processes in Table~\ref{tab:limits} range from $\sim 80$~$\mu$b to $8$~mb. In Fig.~\ref{fig:cross_section_limits} we show the $95\%$ confidence limits on the new physics cross sections as a function of the number of events for a proton at $10^9$~GeV where no systematic uncertainty on the background is assumed.

\section{Conclusions}
\label{sec:conclusions}
Ultra-high-energy cosmic rays interacting with the atoms in the atmosphere are natural high-energy hadron colliders. In comparison with the LHC the event rate recorded through the fly eye at Auger is much smaller. However, the collision energies recorded reach beyond $\mathcal{O}(100)$ TeV. Thus, Auger might become more sensitive than the LHC for new physics scenarios that are realised at energies outside the kinematic reach of the LHC, and for cross sections that are comparable with QCD interactions. Examples of such scenarios would be potentially unsuppressed sphaleron production or a strongly coupled dark sector.
We find that it is possible to set a model-independent limit on the cross sections of such new physics processes by considering their effects on cosmic-ray air showers via the observables $\rho_\mu$ and $X_{\mathrm{max}}$. Using multi-variate data analysis techniques, a strong separation between signal and QCD background interactions can be achieved. However, based on our classification approach, this is only possible for proton primary particles as the effect is washed out for heavier primaries.

\acknowledgments
We would like to thank Mikael Chala and Christoph Englert for interesting discussions, and Stephen Webster for help with Herwig.

%%%%%%%%%%%%%%%%%%%%%%%%%%%%%%%%%%%%%%%%%%%%%%%%%%%%%
%  bibliography
%%%%%%%%%%%%%%%%%%%%%%%%%%%%%%%%%%%%%%%%%%%%%%%%%%%%%
  
\bibliographystyle{eplbib}
\bibliography{references}

\end{document}